\documentclass[aps,twocolumn,showpacs,preprintnumbers,amsmath,amssymb,floatfix]{revtex4}

\usepackage{graphicx}
\usepackage{dcolumn}
\usepackage{bm}

\newcommand{\kpe}{\mathbf{k}\!\cdot\!\mathbf{p}\,}

\newcommand{\etal}{\textsl{et al.}}
\newcommand{\tn}[1]{\textnormal{#1}}

\newcommand{\ingan}{\ensuremath{\tn{In}_\tn{x}\tn{Ga}_\tn{1-x}\tn{N}}}

\begin{document}


\title{GaN/AlN Quantum Dots for Single Qubit Emitters}

\author{M.~Winkelnkemper\footnote{Also at: Fritz-Haber-Institut der Max-Planck-Gesellschaft, Faradayweg 4-6, D-14195 Berlin, Germany}} 
 \email{Momme@sol.Physik.TU-Berlin.de}
\author{R.~Seguin}
\author{S.~Rodt}
\author{A.~Hoffmann}
\author{D.~Bimberg}
\affiliation{Institut f\"ur Festk\"orperphysik, Technische
Universit\"at Berlin, Hardenbergstra{\ss}e 36, D-10623 Berlin, Germany}

\date{\today}

\begin{abstract}
We study theoretically the electronic properties of $c$-plane GaN/AlN quantum dots (QDs) with focus on their potential as sources of single polarized photons for future quantum communication systems. Within the framework of eight-band $\kpe$ theory we calculate the optical interband transitions of the QDs and their polarization properties. We show that an anisotropy of the QD confinement potential in the basal plane (e.g.\ QD elongation or strain anisotropy) leads to a pronounced linear polarization of the ground state and excited state transitions. An externally applied uniaxial stress can be used to either induce a linear polarization of the ground-state transition for emission of single polarized photons or even to compensate the polarization induced by the structural elongation.\\
(accepted at \emph{IOP Journal of Physics: Condensed Matter}; \copyright IOP 2008)
\end{abstract}

\maketitle

\section{Introduction}
Electrically triggered sources of single photons are essential building blocks for future quantum communication and quantum computing systems. Devices based on single quantum dots (QDs) are promising candidates for reliable low-cost solutions in this matter.  Single-photon emitters (SPEs) with extraordinary spectral purity have been realized recently using InAs/GaAs QDs \cite{lochmann2006,scholz2007}.  Here, emission with controlled linear polarization is achieved at low temperatures by exploiting the fine-structure splitting (FSS) of  the exciton ground-state. At elevated temperatures, however, this splitting, of up to 500\,$\mu$eV for large QDs \cite{seguin2005}, is smaller than the homogeneous line width, inhibiting a spectral separation of the co-polarized line pair. Nitride-based QDs present important alternatives for single-photon emission. In $\mu$-photoluminescence (PL) and cathodoluminescence (CL) spectroscopy narrow emission lines from single InGaN/GaN \cite{moriwaki2000,oliver2003,seguin2004,bartel2004,schoemig2004,rice2005,sebald2006,winkelnkemper2007} and GaN/AlN QDs grown on the $c$-plane \cite{kako2004,bardoux2006,kako2006,bardoux2008,simeonov2008} and $a$-plane \cite{rol2007} have been demonstrated. Also zinc-blende GaN/AlN QDs are under investigation \cite{simon2003,lagarde2008}. In the future the spectral range of nitride QDs can be extended to the long-wavelength range for fibre-optical communication employing InN/GaN \cite{ruffenach2007}.  Optically pumped single-photon emission from single GaN/AlN QDs has already been demonstrated \cite{kako2006}. For InGaN/GaN QDs a polarization mechanism based on valence-band (VB) mixing effects has been discovered recently \cite{winkelnkemper2007,winkelnkemper2008} that leads to a separation of differently polarized lines more than an order of magnitude larger than the FSS in InAs/GaAs QDs. Similar polarization effects have been reported for $c$-plane GaN/AlN QDs shortly after \cite{moshe2008,bardoux2008}.   

In this paper, we show that VB mixing effects can be exploited to control the polarization of photons emitted by single GaN/AlN QDs.  First, the polarization mechanism observed for InGaN/GaN QDs will be revisited in Sec.~\ref{sec_ingan}. Then, we will turn our attention to the GaN/AlN system.  We calculate the polarization of the optical interband transitions for GaN/AlN QDs using strain-dependent eight-band $\kpe$ theory (Sec.~\ref{sec_structure}).  The model accounts for piezo- and pyroelectric effects and is implemented for arbitrarily shaped QDs. It is described in more detail in appendix~\ref{sec_method}. 

We show that an asymmetric strain field in the growth plane inside the QDs leads to a linear polarization of the confined A- and B-type excitonic states in orthogonal directions. The in-plane strain anisotropy can originate from different sources, such as QD elongation, inhomogeneous composition profiles, etc.  As examples a slight elongation of the QDs (Sec.~\ref{sec_gan_elong}) and, as an ex-situ approach, the application of external uniaxial stress (Sec.~\ref{sec_gan_press}) are considered in detail.

\section{\label{sec_ingan}Polarization Effects in InGaN/GaN Quantum Dots}
We have shown recently that emission lines from single InGaN/GaN QDs are systematically linearly polarized in the orthogonal crystal directions [$11\overline{2}0$] and [$\overline{1}100$]---a symmetry that is non-native to hexagonal crystals \cite{winkelnkemper2007, winkelnkemper2008}.   The polarization of the emission lines has been attributed to the character of the hole state [formed by either the $A$ or $B$ valence band (VB)] involved in the recombination process and a slight elongation of the QDs.

 The \ingan/GaN samples were grown in the wurtzite phase by metal-organic chemical vapor deposition on Si(111) substrates. The InGaN layers were grown at 800\,$^{\circ}$C with a nominal thickness of 2\,nm using trimethylgallium, trimethylindium, and ammonia as precursors. The QDs are formed by alloy fluctuations  within the InGaN-layer \cite{seguin2004}.  The sample has been investigated with single-QD CL employing Pt shadow masks to probe a few QDs only.  Further discrimination of the single-QD lines was possible using their temporal jitter: All lines originating from the same QD show the same characteristic jitter pattern.  Line groups of up to five narrow lines with the same jitter pattern could be identified \cite{winkelnkemper2007}.  An example of such a single-QD spectrum is shown in Fig.~1. All emission lines show a pronounced linear polarization. The polarization directions scatter around [$\overline{1}100$] and [$11\overline{2}0$]. Both directions were found in each line group \cite{winkelnkemper2007}.  
\begin{figure}
\centering
\includegraphics[clip,width=\columnwidth]{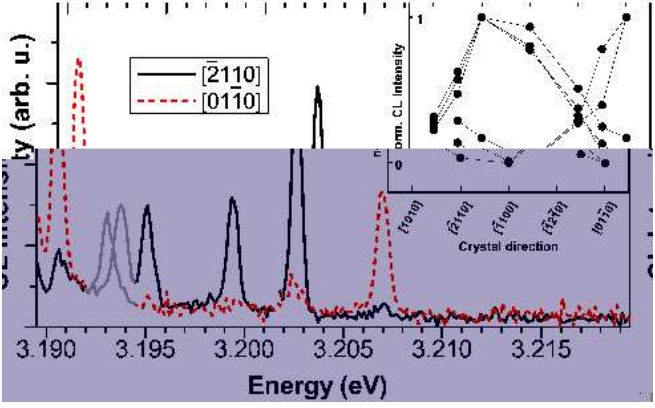}
\caption{\label{fig_ingan_spec}(Color online) CL spectrum of a single InGaN/GaN QD. Grey lines belong to a different QD. The inset shows the evolution of the intensities of the five lives as a function of the polarization angle.}
\end{figure}
The emission patterns differ significantly from the well-known polarized emission spectra of II/VI and III/V QDs with cubic crystal structure. Exchange-interaction induced FSS of the exciton (see, e.g., \cite{bayer2002,kulakovskii1999,seguin2005}) or other excitonic complexes has not yet been reported for nitride QDs. 

In contrast to InAs/GaAs QDs or GaN/AlN QDs, which grow in the strain-induced Stranski-Krastanov or Vollmer-Weber growth modes, the \ingan/GaN QDs form within the \ingan\ layer due to composition fluctuations within the layer. For a theoretical description these QDs have been modeled as ellipsoids with linearly graded indium concentration \cite{winkelnkemper2006,winkelnkemper2007,winkelnkemper2008}. The maximum indium concentration $x_\tn{c}$ at the QDs' centers  is $50$\,\%, the minimum concentration $x_\tn{e}$ at the QDs' borders  is $5$\,\%. The QDs are embedded in an $\tn{In}_{0.05}\tn{Ga}_{0.95}\tn{N}$ quantum well with a height of $2$\,nm. The QDs have a height of $d_z=2$\,nm and lateral extensions of $d_{x/y}=4.6-5.8$\,nm. Henceforth $x$ will denote the direction of the QDs long axis. Further details on the calculations can be found in appendix~\ref{sec_method}.
\begin{figure}
\centering
\includegraphics[clip,width=\columnwidth]{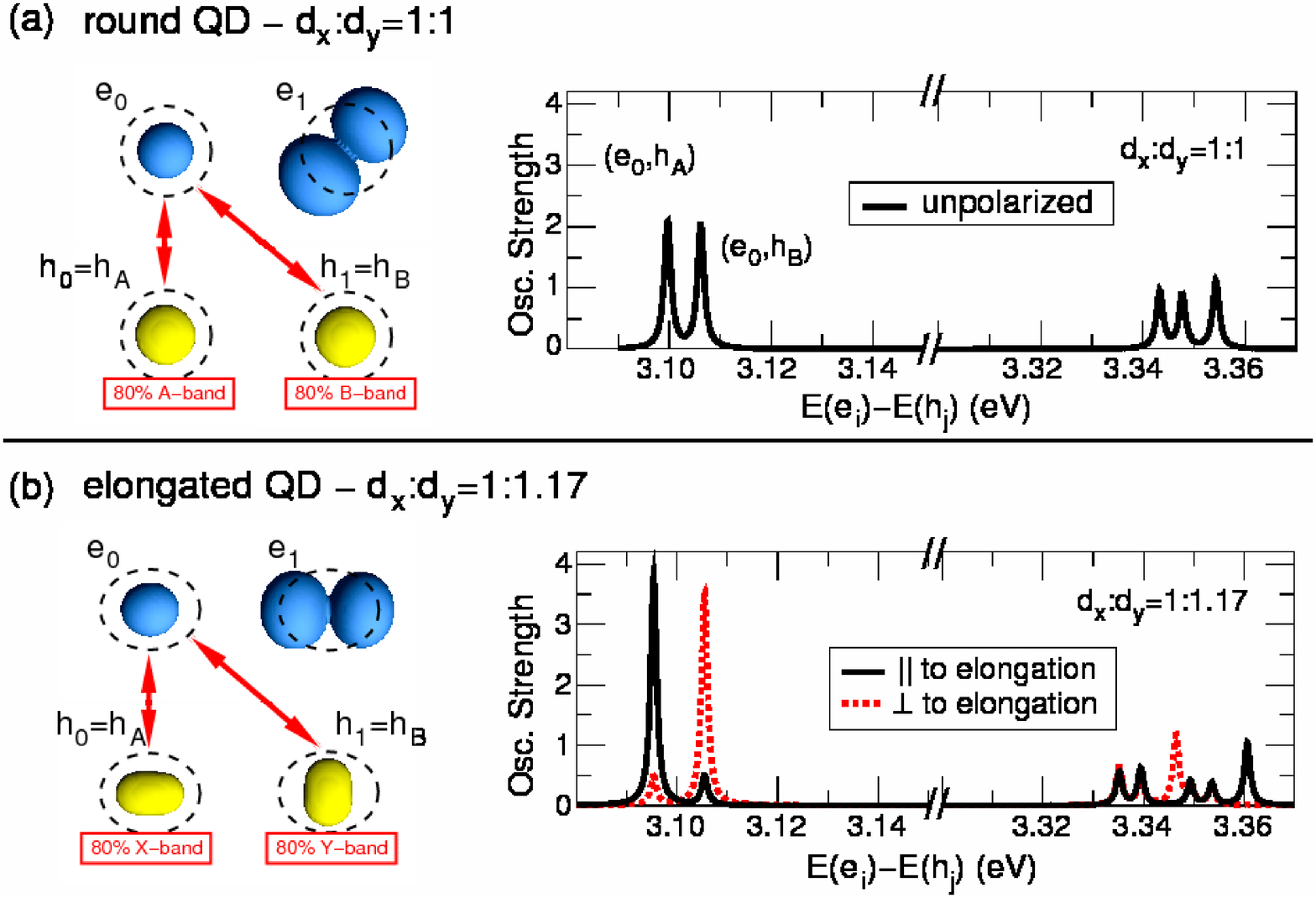}
\caption{\label{fig_ingan_states}(Color online) Confined single-particle electron (blue) and hole (yellow) states in a round (a) and an elongated (b) QD. The corresponding interband transitions are shown on the right-hand side. For the round QD, the $A$- and $B$-band transitions emit unpolarized light (with $k\parallel C$), while the respective transitions in the elongated QD are polarized parallel and perpendicular to the long axis of the QD. The transitions between 3.33 and 3.36\,eV originate from higher excited electron and hole levels.}
\end{figure}

The emission lines can be attributed to transitions involving either the confined hole ground state ($h_0\equiv h_A$), which is predominantly formed by the $A$ band, or the first excited hole state ($h_1\equiv h_B$), which is predominantly formed by the $B$ band.  If a QD is slightly elongated, transitions involving $h_A$ are polarized parallel to the elongation and transitions involving $h_B$ perpendicular to it \cite{winkelnkemper2007,winkelnkemper2008}.
The single-particle electron and hole orbitals are depicted in Fig.~2 for a circular QD (a) and a slightly elongated QD (b).  The electron ground-state ($e_0$) envelope functions have $s$-like symmetry; the envelope functions of the first excited electron state ($e_{1}$) is $p$-like.  In contrast, the first two hole states, $h_A$ and $h_B$, both have $s$-like envelope functions.  Both states have sizable oscillator strengths with the electron ground state, but behave differently if the strain field within the QD is anisotropic in the basal plane, e.g., due to an elongated QD shape: $h_A$ aligns parallel to the QD's long axis, $h_B$ perpendicular to it.  An analysis of the projections of both hole states on the $\kpe$ basis states $P_x$ and $P_y$  yields that the $P_x$ ($P_y$) projection of $h_A$ ($h_B$) increases with increasing QD elongation, while the $P_y$ ($P_x$) projection decreases \cite{winkelnkemper2008}. Consequently, the optical transition between $h_A$ ($h_B$) and $e_0$ is linearly polarized parallel (perpendicular) to the long axis of the QD.
\begin{figure}
\centering
\includegraphics[clip,width=\columnwidth]{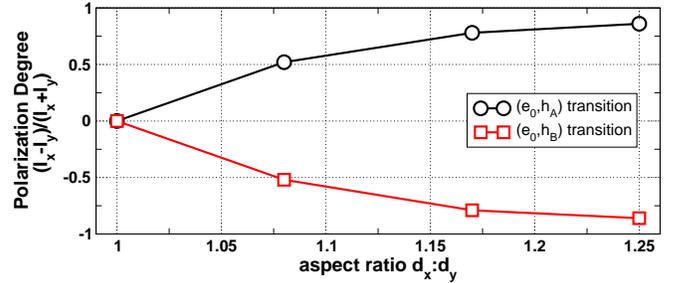}
\caption{\label{fig_ingan_poldeg}(Color online) Degree of linear polarization of the $e_0$-$h_0$ (black line and circles) and $e_0$-$h_B$ (red line and squares) transitions as a function of the in-plane elongation of \ingan/GaN QDs.}
\end{figure}
The polarization increases with increasing QD elongation (Fig.~\ref{fig_ingan_poldeg}).
\section{\label{sec_structure}GaN/AlN Quantum Dots}
\begin{figure}
\centering
\includegraphics[clip,width=\columnwidth]{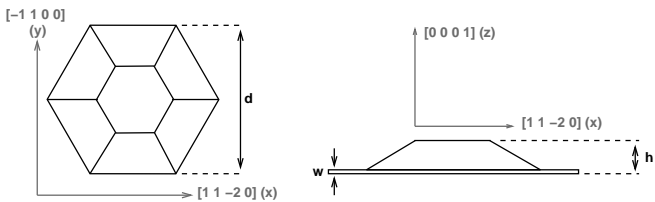}
\caption{\label{fig:cplane_struc}Model structure of $c$-plane GaN/AlN QDs: Truncated hexagonal pyramid with height $h$ between $1.5$\,nm and $3.5$\,nm. The lateral diameter $d$ is determined by the dot's aspect ratio $h$:$d=1$:$10$ ($1$:$5$). The wetting-layer thickness is  $w=0.25$\,nm.}
\end{figure}
\begin{figure}
\includegraphics[clip,width=\columnwidth]{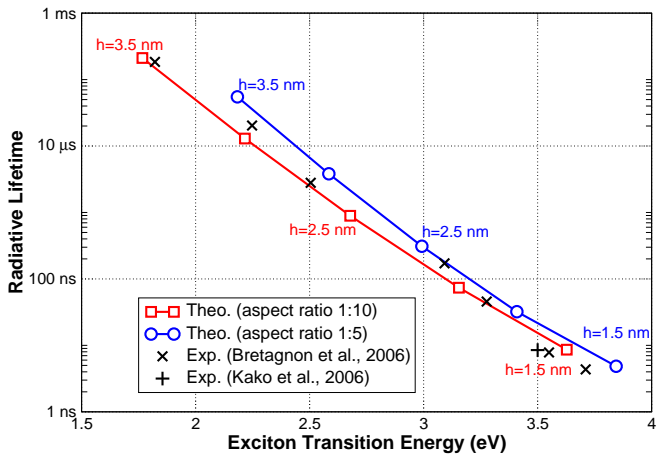}
\caption{\label{fig:cplane_tau} (Color online) Radiative lifetimes of confined excitons in $c$-plane GaN/AlN QDs: \emph{Red lines and squares:} Calculated, QDs with an aspect ratio of $h$:$d=1$:$10$ . \emph{Blue line and circles:}  Calculated, QDs with an aspect ratio of $h$:$d=1$:$5$. \emph{Black crosses:} Experimental values from Bretagnon \etal\cite{bretagnon2006}. \emph{Black plus:}  Experimental value from Kako \etal\cite{kako2006}.} 
\end{figure}

In contrast to the \ingan/GaN QDs discussed in Sec.~\ref{sec_ingan}, GaN/AlN QDs grow strain-induced in the Stranski-Krastanov or Vollmer-Weber growth mode. Also their typical size and shape is known with greater accuracy. Experimental reports on the structural properties of $c$-plane GaN QDs (e.g., Refs.~\onlinecite{daudin1997,widmann1998,kako2004,kako2006,hoshino2004}) agree on the shape of the QDs, which is a truncated hexagonal pyramid. The reported heights ($h$) scatter between $1.3$\,nm and $5$\,nm. The aspect ratios ($h$:$d$, where $d$ denotes the lateral diameter) in most reports are in the range of 1:5 to 1:10. The model structure derived from these reports is depicted in Fig.~\ref{fig:cplane_struc}.   The model QDs have an aspect ratio of either $h$:$d=1$:$10$ or  $1$:$5$. The height of the QDs is varied between $1.5$\,nm and $3.5$\,nm in steps of $0.5$\,nm.  The thickness of the wetting layer is assumed to be $w=0.25$\,nm. This set of model QDs covers the major part of experimentally reported QD structures and yields excitonic transition energies and radiative lifetimes in good agreement with experimental values \cite{bretagnon2006,kako2006} (Fig.~\ref{fig:cplane_tau}). 
The huge built-in piezo- and pyroelectric fields within GaN/AlN QDs strongly affect the emission energies and radiative lifetimes of localized excitons within the QDs via the quantum-confined Stark effect (QCSE) \cite{widmann1998,andreev2001,bretagnon2003,bretagnon2006}.  In the center of the QD these fields are as large as $9.5$\,MV/cm for QDs with an aspect ratio of $1$:$10$ and $8.0$\,MV/cm for QDs with an aspect ratio of $1$:$5$. Depending on the size of the QDs the radiative lifetimes range from a few nanoseconds for small QDs up to as long as 100\,{$\mu$}s for large QDs. Radiative lifetimes in this range would limit cut-off frequencies of devices to a maximum of a few kHz. Thus, with the fields present, only small $c$-plane QDs, i.e. those with high transition energies, are good candidates for fast single-photon emitters (SPEs).  
\subsection{Single-particle Energy Levels}
  The bound hole states in GaN/AlN QDs are---as in the case of \ingan/GaN QDs---formed predominantly by the $A$ and $B$ band. $C$-band contributions are small because the biaxial strain within the QDs shifts this band to much lower energies \cite{winkelnkemper2006,winkelnkemper2007}.  As a first approximation, for each band we expect to find a ground state with an $s$-shaped envelope function, which is only spin degenerate. The $p$ shell consists of two states and the $d$ shell of three. Due to the different parities of the bulk conduction and valence bands, the electron and hole states have a finite optical matrix element if their envelope functions have \emph{the same} parity, i.e.\ the allowed transition channels are $s$-$s$, $p$-$p$, $s$-$d$, etc. Each transition channel exists twice, once for the $A$-type holes and once for the $B$-type holes.
\begin{figure}
\includegraphics[clip,width=\columnwidth]{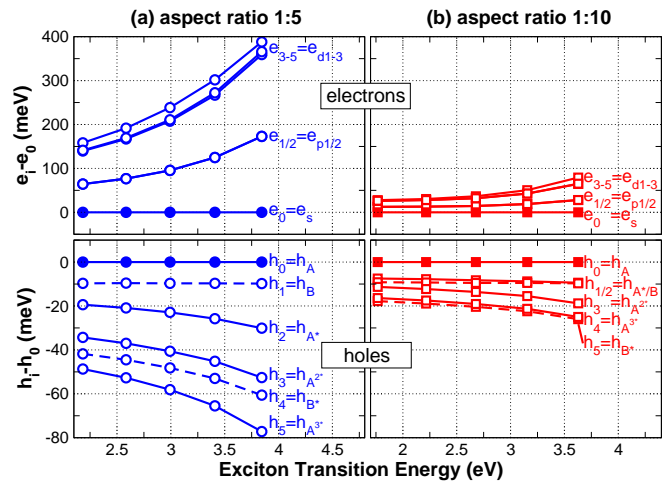}
\caption{\label{fig:cplane_single} (Color online) Energies of the single-particle electron and hole states. The electron states are labeled according to the symmetry of their envelope functions ($s$, $p$, and $d$ shell). Hole-state labels are given according to the valence band from which the hole states are predominantly formed.} 
\end{figure}
Figure~\ref{fig:cplane_single} shows the calculated single-particle electron and hole energy levels of all QDs considered in this work including the six energetically lowest (highest) electron (hole) states. The electron states are all formed predominantly by the conduction band ($\approx 95$\,\%).  Therefore, the $s$, $p$, and $d$ shells can be clearly distinguished.  The hole spectra, however, are more complex, because holes states are not formed by either the $A$ \emph{or} $B$ band, but by a mixture of both bands and even a small $C$-band contribution. Still we can characterize each hole state by the band that contributes the most to it (see labels in Fig.~\ref{fig:cplane_single}). For QDs with an aspect ratio of $1$:$5$ [Fig.~\ref{fig:cplane_single}(a)] the $A$-band $s$-state ($h_0\equiv h_A$; $\approx95$\,\% $A$-band projection) and the $B$-band $s$-state ($h_1\equiv h_B$; $\approx90$\,\% $B$-band projection) are energetically well separated from the excited hole states. Both have an unambiguously $s$-shaped envelope function (not shown here). The splitting between both states ($\approx 9-10$\,meV) does not increase for smaller QDs, but is constant.  It corresponds to the energy separation between $A$ and $B$ band in strained GaN. The higher excited hole states can not be unambiguously assigned to $p$- or $d$-like orbitals. Please note, that although they have been labeled according to the major band contributions, this contribution sometimes does not exceed 50\,\%.  The QDs with aspect ratio $1$:$10$ [Fig.~\ref{fig:cplane_single}(b)] exhibit a significantly smaller excited-states splitting for electrons and holes due to the weaker lateral confinement. An exception is the splitting between $h_A$ and $h_B$, which is largely independent from the QD dimensions as discussed before.

\subsection{\label{sec_gan_elong}Optical Transitions and their Polarization}
\begin{figure}
\centering
\includegraphics[clip,width=\columnwidth]{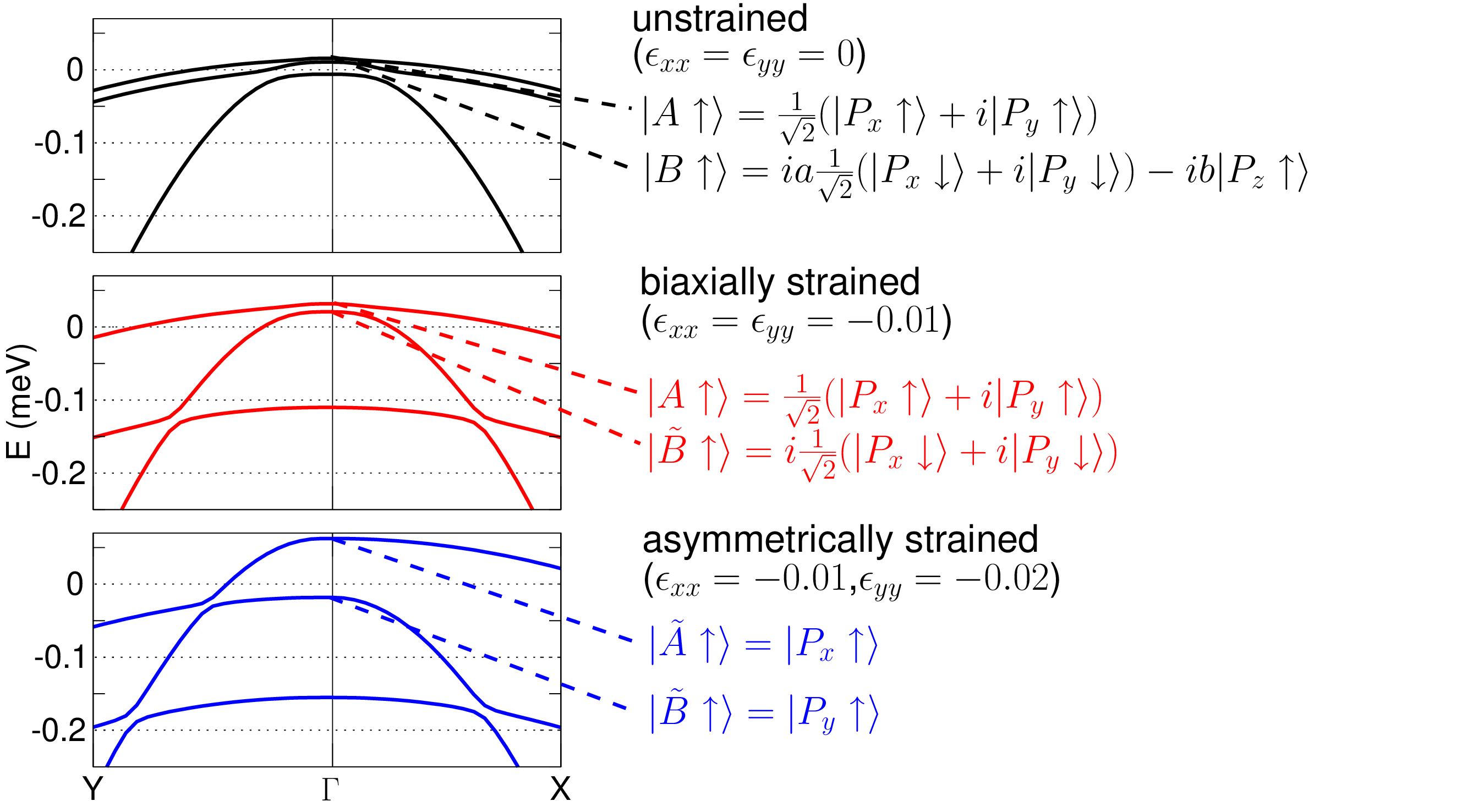}
\caption{\label{fig_bands} (Color online) Valence band structure of GaN in the vicinity of $\Gamma$. \emph{Upper panel:} unstrained. \emph{center:} under symmetric biaxial strain in the basal plane, typical for $c$-plane QDs. \emph{Lower:} With an additional strain anisotropy in the basal plane, e.g., in asymmetric QDs. The splitting between the two top bands increases and their characters change from $A$- and $B$-like to $P_x$-like and $P_y$-like.} 
\end{figure}
\begin{figure}
\centering
\includegraphics[clip,width=\columnwidth]{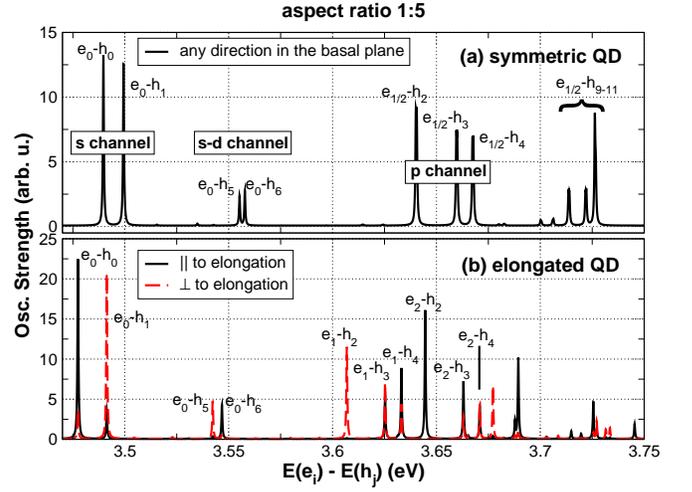}
\caption{\label{fig:cplane_transitions02} (Color online) Oscillator strength between the single-particle electron and hole state for a QD with an aspect ratio of $h$:$d=1:5$ and a height of $h=2.0$\,nm. Shown are all transitions involving one of the first six electron levels ($e_i$) and one of the first twelve hole levels ($h_j$), but only up to an energy difference $e_i-h_j$ of $3.75$\,eV. (a) For a symmetric QD. (b) For a QD with a 10\,\% in-plane elongation along [$11\overline{2}0$].} 
\end{figure}
In bulk GaN an anisotropy of the strain field in the basal plane changes the character of the band-edge states from $A$ ($B$) type to $P_\tn{x}$ ($P_\tn{y}$), if the stress is larger along $y$ (see Fig.~\ref{fig_bands}). The optical matrix elements involving the confined $A$- or $B$-band hole states in the QDs respond differently to an asymmetry of the confinement potential in the basal plane: They become linearly polarized in orthogonal directions. Figure~\ref{fig:cplane_transitions02}(a) shows the oscillator strengths between the six lowest electron levels $e_i$ and the twelve highest hole levels $h_j$ (up to an energy difference $e_i-h_j$ of $3.75$\,eV) for the QD with $h$:$d=1$:$5$ and $h=2.0$\,nm.  Figure~\ref{fig:cplane_transitions02}(b) shows the respective spectrum for a similar QD with an in-plane elongation of 10\,\% along [$11\overline{2}0$].  Almost all transitions of the elongated dot show a pronounced linear polarization either parallel to the elongation or perpendicular to it. The transition between the electron ground state $e_0$ and $h_A$ ($h_B$) is linearly polarized along (perpendicular to) the long axis of the QD. Both transitions together can be regarded as $s$-channel. The wave functions of the higher excited hole states are build of sizeable contributions from more than one valence band. Therefore not all of these states can be unambiguously assigned to $p$- or $d$-like symmetry. The $p$- and $d$-shell labels in Fig.~\ref{fig:cplane_transitions02} are given according to the main contribution.   The QDs with aspect ratio $h$:$d=1$:$10$ show the same polarization as those with $1$:$5$ albeit less pronounced. The strain anisotropy that is induced by the elongation affects the confined states less in QDs with larger diameter, because the main parts of the wave functions are located in the center of the dot.   The energetic separation between the orthogonally polarized ($e_0$,$h_A$) and ($e_0$,$h_B$) transitions in the $s$-channel is independent of the size of the QDs or their vertical aspect ratio. For (in-plane) symmetric QDs it is about $10$\,meV and increases for anisotropic QDs. This large separation between the orthogonally polarized lines should enable a spectral separation of both lines even at elevated temperatures. Thus, the well-defined polarization of the ground-state transition can be exploited for polarization control in future SPEs.

\subsection{\label{sec_gan_press}Control of the Polarization by Uniaxial Stress}
\begin{figure}
\centering
\includegraphics[clip,width=\columnwidth]{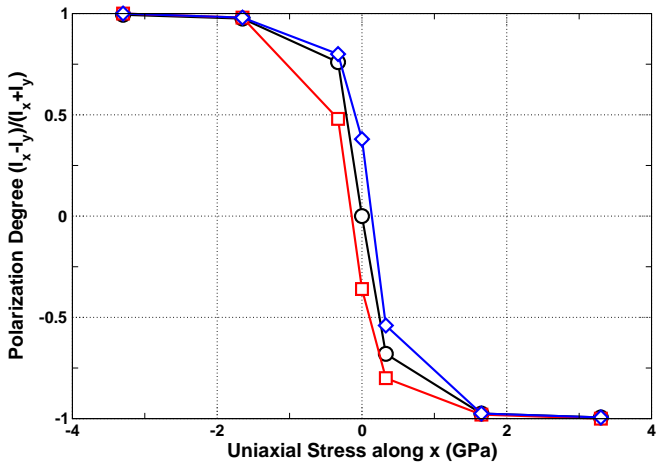}
\caption{\label{fig:cplane_presspol} (Color online) Polarization degree of the excitonic ground-state transition as a function of uniaxial stress along $x$ ($[11\overline{2}0]$), relaxation of the sample in $y$ and $z$ direction according to the laws of continuum mechanics has been included. The calculations have been performed for three different QDs with an aspect ratio of $h$:$d=1$:$10$ and a height of $h=2.0$\,nm. \emph{Black lines and circles:} Symmetrical QD.  \emph{Blue lines and diamonds:} QDs with 10\,\% elongation along $x$. \emph{Red line and squares:} QDs with 10\,\% elongation along $y$.} 
\end{figure}
Since the linear polarization of the transitions is a result of the strain anisotropy, it can be induced by uniaxial stress, without any structural anisotropy of the dot. The effect of uniaxial stress on the polarization of the ground-state transition of the QD with aspect ratio $1$:$10$ and height $2$\,nm is shown in Fig.~\ref{fig:cplane_presspol} (black line and circles). A pronounced polarization is found at stress levels that can easily be induced by anisotropic strain relaxation in epitaxial heterostructures (due to the formation of cracks or defects) or by externally applied stress. Thus, it is possible to control the polarization of the emitted photons.  Figure~\ref{fig:cplane_presspol} also shows similar calculations for elongated QDs (blue lines and diamonds; red lines and squares). The polarization induced by the structure can be easily enhanced or even inverted by the external stress. 

\section{Summary}
We have theoretically investigated the spectroscopic properties of $c$-plane GaN/AlN QDs with a special emphasis on their suitability as sources of single polarized photons.  We have shown that the linear polarization of the interband transitions is effectively controlled by an anisotropic strain field within the QDs. Transitions involving either $A$- or $B$-type hole states are polarized in orthogonal directions. The separation of the $A$-type ground-state and the orthogonally polarized $B$-type first excited state is $\approx 10$\,meV and largely independent from the QD size and shape. 

A sufficient anisotropy can be induced either by a structural elongation of the QDs or by an externally applied uniaxial stress. An in-plane elongation of the QDs of only 10\% leads to a polarization degree of the excitonic ground-state transition of up to $6$:$1$ depending on the other structural parameters of the QDs.  An externally applied uniaxial stress of about 300\,MPa leads to a polarization degree of more than 5:1; larger stress results in a complete polarization of the emission. Moreover, a polarization resulting from structural elongation of the QDs can be compensated by external stress. This effect could be exploited for future devices, in particular to achieve a well-defined polarization in QD-based single-photon emitters.

\section*{Acknowledgements}
The authors would like to thank C.~Kindel and A.~Schliwa for fruitful discussions.  This work was supported by Deutsche Forschungsgemeinschaft in the framework of Sfb 787 and by SANDiE Network of Excellence of the European Commission (No.\ NMP4-CT-2004-500101). Parts of the calculations were performed on an IBM p690 supercomputer at HLRN Berlin/Hannover with project No.~bep00014.
\appendix
\section{\label{sec_method}Calculation Method}
For the electronic structure calculations we use a three-dimensional eight-band $\mathrm{\textbf{k} \cdot \textbf{p}}$ model implemented on a finite differences grid.  The model accounts for strain effects, piezoelectric and pyroelectric polarization, spin-orbit and crystal-field splitting, and coupling between the valence bands (VBs) and the conduction band (CB).  The $\kpe$ method is based on the one originally introduced by Stier \textsl{et al.} \cite{stier1999}. The implementation for wurtzite crystals is described in detail in Ref.~\onlinecite{winkelnkemper2006}.  Details of the calculation of radiative lifetimes are given in Ref.~\onlinecite{winkelnkemper2008_2}.
Few-particle corrections have been included using a self-consistent Hartree (mean-field) scheme \cite{winkelnkemper2006}.   For the \ingan/GaN system material parameters from Ref.~\onlinecite{rinke2006} have been included as described in Ref.~\onlinecite{winkelnkemper2007}. 
Most of the material parameter used for the GaN/AlN system are based on the recommendations of Vurgaftman and Meyer \cite{vurgaftman2003}.  For the parameters $m_\tn{e}^\tn{i}$, $A_\tn{i}$, and $E_\tn{P}^\tn{i}$ we use recently developed band-dispersion parameters \cite{rinke2006,rinke2008} derived from accurate quasiparticle energy calculations \cite{rinke2005}. For the valence-band offset between GaN and AlN we use the value of $0.8$\,eV determined by Wei and Zunger \cite{wei1996}.   As already detailed in Ref.~\onlinecite{winkelnkemper2006}, we neglect the anisotropy of the static dielectric constants $\epsilon_\tn{r}$ and use the mean value for all direction. For the refractive index of AlN we use the Sellmeier-formula derived by Antoine-Vincent \etal \cite{antoine2003}.

\end{document}